\documentclass{iaus}
\usepackage{graphicx}

\title[Analysis of XX Leo] 
{The BVRI Light Curves and Period Analysis of the Beta Lyrae
System XX Leonis}

\author[Zasche et al.]   
{P. Zasche$^1$, 
 P. Svoboda $^2$ \and M. Wolf $^1$}

\affiliation{
$^1$ Astronomical Institute, Faculty of Mathematics
and Physic, Charles University, \break V Hole\v{s}ovi\v{c}k\'{a}ch
2, CZ - 180 00 Praha 8, Czech Republic,
\break email: petr.zasche@email.cz \\[\affilskip]
$^2$ Private observatory, V\'{y}pustky 5, CZ - 614 00 Brno, Czech
Republic}

\pubyear{2006}
\volume{240}  
\pagerange{119--126}
\date{?? and in revised form ??}
\jname{Proceedings Title IAU Symposium}
\editors{A.C. Editor, B.D. Editor \& C.E. Editor, eds.}
\begin{document}

\maketitle

\begin{abstract}
The contact eclipsing binary system XX Leonis (p = 0.97 days, Sp
A9) has been analysed using the PHOEBE programme, based on the
Wilson-Devinney code. The BVRI light curves were obtained during
spring 2006 using the 20-cm telescope and ST-7 CCD detector. The
effective temperature of the primary component determined from the
photometric analysis is $T=(7342\pm14)\,K$, the inclination of the
orbit is $i=(84.83\pm0.29)^{\circ}$ and the photometric mass ratio
$q=(0.40\pm0.01)$. Also the third body hypothesis was suggested,
based on the period analysis using 57 minimum times and resulting
the period of the third body $p_3=(59.66\pm0.05)\,yr$, amplitude
$A=(0.036\pm0.028)\,day$ and zero eccentricity which gives the
minimum mass $M_{3,min}=(0.91\pm0.01)\,M_{\odot}$.
\end{abstract}

\firstsection 

\section{Introduction}

The binary system XX Leonis (SV* P 3370, AN 355.1934, BD+14 2177)
was discovered to be a variable by \cite{Tsesevich1954}. It is an
eclipsing binary star of $11^{th}$ magnitude, the period is close
to one day and the depth of primary minimum is about 0.5 mag. The
spectral type of the primary is A9V and of secondary K0. It was
included in a catalogue of apsidal motion binaries by
\cite{Heged1988}, but this hypothesis was rejected.

\section{Light curve analysis}

The preliminary solution of the light curve (using Washington
filter system) computed by \cite{Shaw1998} classified XX Leo as a
near-contact binary, but the period analysis indicates constant
period. The second analysis with satisfactory data was performed
by Stark et al. in 2000 but with only R and V light curves and
also without a satisfactory light curve solution. So the masses of
the components are still not very convincing. We assume here
$M_1+M_2=2.41\,M_{\odot}$, according to the derived spectral types
from the previous analysis.

\begin{figure}
  \scalebox{0.70}{\includegraphics*[14mm,100mm][210mm,183mm]{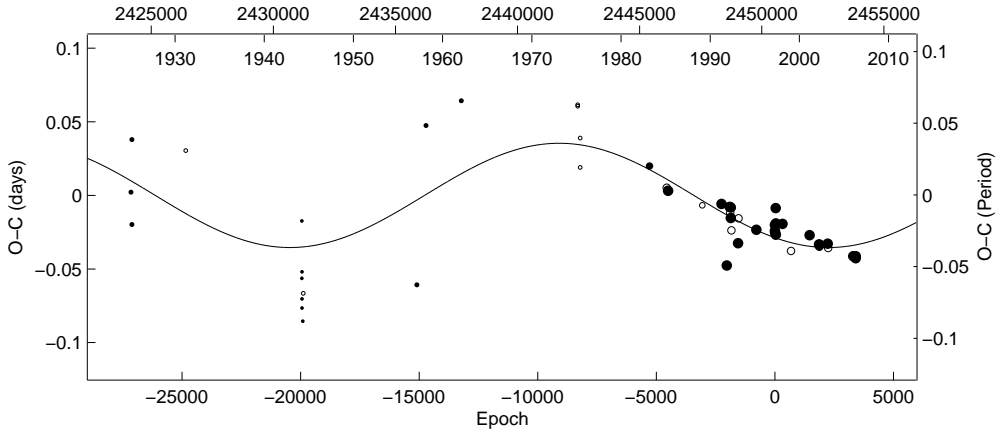}}
  \caption{O-C diagram of XX Leo.}\label{fig:LITE}
\end{figure}

The new light elements derived from our period analysis using 57
minimum times (see Figure \ref{fig:LITE}) for XX Leo are:
\begin{eqnarray}
 \mathrm{Min{\,}I} &=& \mathrm{HJD} \;\;24\;50540.6831 + 0.\!\!^{\rm d}971135899 \cdot {\rm E}.\\[-0.5mm]
 & & \hspace{18.5mm} \pm 0.\!0076 \pm 0.\!\!^{\rm d}000000473
\nonumber
\end{eqnarray}

\begin{figure}
  \scalebox{0.85}{\includegraphics*[24mm,170mm][181mm,290mm]{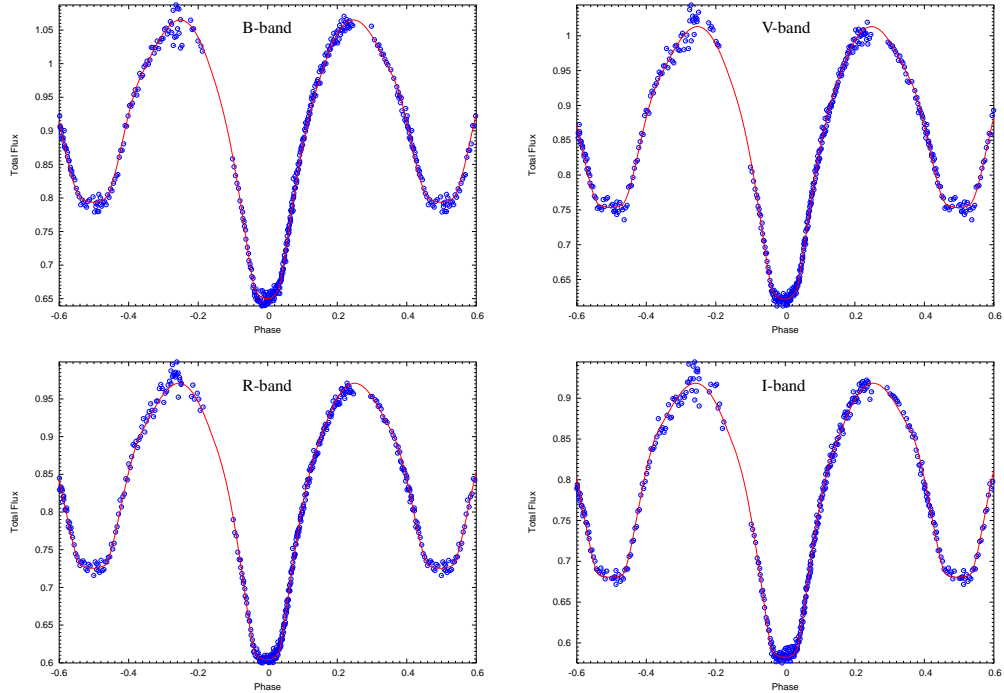}}
  \caption{Light curves of XX Leo measured through Bessell filters B,V,R and I during
  spring 2006. Altogether 1495 data points (with no weightening scheme) and
  also the theoretical curves are plotted. On the x-axis is the phase
  according to the ephemeris above and on the y-axis is the relative total
  flux from the system.}\label{fig:LC}
\end{figure}

\noindent We have measured the light curves using standard Bessell
B,V,R,and I filter system with the 20-cm telescope and ST-7 CCD
detector. These measurements were done in private observatory in
Brno, Czech Republic, during 20 nights in spring 2006. In the
Figure~\ref{fig:LC} you can see the theoretical curves with the
individual data points. These measurements were analysed using the
PHOEBE programme (see \cite{Prsa2005}), based on the
Wilson-Devinney code (see eg. \cite{Wilson1971}). The final values
for both components for the temperatures ($T_i$), mass ratio
($q$), inclination ($i$), relative luminosities ($L_1, L_2, l_3$),
limb-darkening ($x_i$, interpolated from the Van-Hamme's tables),
albedo ($A_i$) and gravity darkening ($g_i$) coefficients and
synchronicity parameters ($F_i$), respectively, are in the Table
\ref{tab:par}. We have used the "double contact binary" mode
during the computation, because of the best numerical result, the
best agreement with the previous results (primary and secondary
spectral type A9 and K0, respectively) and also the period
analysis. Regrettably we have no RV curve, so we cannot estimate
another parameters more precisely, for example the masses and
dimensions of the system. The mass ratio computed only from the
photometry is not very reliable. Spectroscopy of XX Leo would help
us to estimate these parameters.

From the mass of the potential third body (about $0.9 M_{\odot}$,
from the period analysis) we could conclude that this star on the
main sequence should be G9 spectral type. Also from the light
curve analysis we can see that this body has almost the same
luminosity as the secondary component, so both approaches lead to
the same result.

\section{Period analysis}

The previous period analysis of XX Leo was performed by
\cite{Stark2000}, but no significant variation was found because
of small data set. Also in the paper by \cite{Srivastava1994} was
found that the period of the binary is constant. We have used 57
minimum times covering more than 80 years. In the Figure
\ref{fig:LITE}, there are individual minimum times marked as dots
and circles for the primary and secondary minimum, respectively.
We can see no evident displacement of secondary minima, so the
apsidal motion could be ruled out. Bigger the symbol, bigger the
weight. Also the light-time effect hypothesis curve is shown. The
final results of the fit are in the Table \ref{tab:LITE}. We have
fixed the value of eccentricity to zero because of the best
agreement with the photometric results.

\begin{table}[b!]
 \begin{center}
 \caption{Physical parameters of XX Leo.}
 \label{tab:par}
 \begin{tabular}{cccc}
 \hline
 Parameter &  Value &  Parameter &  Value \\
 \hline 
 $T_1 [K]$      & $7342 \pm 14$    & $x_{1,B}$ & $0.603 \pm 0.009$ \\
 $T_2 [K]$      & $4991 \pm 88$    & $x_{1,V}$ & $0.508 \pm 0.009$ \\
 $M_2/M_1$    & $0.401 \pm 0.002$  & $x_{1,R}$ & $0.408 \pm 0.008$ \\
 $i [^\circ]$ & $84.835 \pm 0.287$ & $x_{1,I}$ & $0.320 \pm 0.010$ \\
 $L_{1,B} $   & $0.891 \pm 0.003$  & $x_{2,B}$ & $0.866 \pm 0.115$ \\
 $L_{1,V} $   & $0.841 \pm 0.002$  & $x_{2,V}$ & $0.724 \pm 0.095$ \\
 $L_{1,R} $   & $0.809 \pm 0.002$  & $x_{2,R}$ & $0.597 \pm 0.079$ \\
 $L_{1,I} $   & $0.783 \pm 0.002$  & $x_{2,I}$ & $0.489 \pm 0.080$ \\
 $L_{2,B} $   & $0.070 \pm 0.002$   & $A_1$ & $0.215 \pm 0.019$ \\
 $L_{2,V} $   & $0.079 \pm 0.002$   & $A_2$ & $0.875 \pm 0.094$ \\
 $L_{2,R} $   & $0.089 \pm 0.001$   & $g_1$ & $0.807 \pm 0.034$ \\
 $L_{2,I} $   & $0.106 \pm 0.001$   & $g_2$ & $0.069 \pm 0.087$ \\
 $l_{3,B} $   & $0.040 \pm 0.003$   & $F_1$ &  0 (fixed)        \\
 $l_{3,V} $   & $0.080 \pm 0.002$   & $F_2$ & $2.249 \pm 0.018$ \\
 $l_{3,R} $   & $0.102 \pm 0.002$   \\
 $l_{3,I} $   & $0.111 \pm 0.002$   \\
 \hline
\end{tabular}
 \end{center}
\end{table}

From the third-body hypothesis also the mass function and minimum
mass of the third body was calculated. The mass function is
$f(m_3)=0.068 M_{\odot}$, so the minimum mass of the third body is
$0.91 \pm 0.02 M_{\odot}$. As we can see, this suggested body
might be a bit more massive than the secondary component, what in
a good agreement according to the luminosities of the individual
components from the light curve analysis. Secondary and tertiary
component seems to be rather similar, what could be interesting
example for spectral disentangling.

\section{Conclusions}\label{sec:concl}

We have performed new analysis of the contact binary of the
Beta-Lyrae type XX Leo. New light curves were measured and
analysed. Also new period study has been done. The hypothesis of
the third body was suggested, according to the third light in the
LC solution and also from the light-time effect hypothesis. Both
hypothesis are in very good agreement and leads to the third body
comparable to the secondary.

\begin{table}\def~{\hphantom{0}}
  \begin{center}
  \caption{Parameters of the third component.}
  \label{tab:LITE}
  \begin{tabular}{cc}\hline
       Parameter &  Value \\\hline
        $p_3 [yr]$ & $59.66 \pm 0.05$ \\
        $A [d]$    & $0.036 \pm 0.028$ \\
        $\omega [^\circ]$ & $19.22 \pm 12.98$ \\
        $e$ & 0 (fixed) \\
        $T_0 [JD]$ & $2437200.62 \pm 216.63$\\\hline
  \end{tabular}
 \end{center}
\end{table}

\begin{acknowledgments}
This research has made use of the SIMBAD database, operated at
CDS, Strasbourg, France, and of NASA's Astrophysics Data System
Bibliographic Services. This investigation was supported by the
Grant Agency of the Czech Republic, grant No. 205/04/2063 and No.
205/06/0217.
\end{acknowledgments}

\end{document}